\documentclass[a4paper,11pt]{article}

\usepackage[utf8]{inputenc}
\usepackage[T1]{fontenc}
\usepackage{lmodern}
\usepackage{geometry}
\usepackage{graphicx}
\usepackage{authblk}
\usepackage{url}
\usepackage{cite}
\usepackage{fancyhdr}
\usepackage[hidelinks]{hyperref}
\usepackage{orcidlink}
\usepackage{amsmath}
\usepackage{xspace}
\usepackage{tikz}
\usepackage{tikz-feynman}
\usetikzlibrary{intersections}
\definecolor{greenMD}{RGB}{115, 155, 58}
\definecolor{purpleMD}{RGB}{140, 100, 197}
\usepackage{subcaption}
\usepackage{cleveref}
\usepackage{comment}
\makeatletter
\g@addto@macro\bfseries{\boldmath}
\makeatother

\newcommand{\Bptotaunu}{\ensuremath{B^+\to \tau^+ \nu_\tau}\xspace}
\newcommand{\Bctotaunu}{\ensuremath{B_c^+\to \tau^+ \nu_\tau}\xspace}
\newcommand{\tautothreepi}{\ensuremath{\tau^+\to \pi^+ \pi^- \pi^+ \bar{\nu}_\tau}\xspace}

\fancypagestyle{firstpage}{
	\fancyhf{}
	\fancyhead[C]{\includegraphics[height=2.2cm]{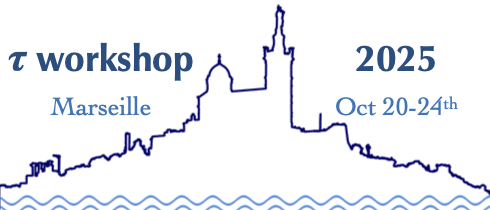}\\ \textit{Proceedings of the 18th International Workshop on Tau Lepton Physics}
	} % <-- put your logo here
}

\title{Looking forward to $B^+\to \tau^+ \nu_\tau$ and $B_c^+\to \tau^+ \nu_\tau$}
\author[1,2,3]{Maria Domenica Galati\,\orcidlink{0000-0002-8716-4440}\thanks{Presenting author.}}
\author[2,3]{Kristof De Bruyn\,\orcidlink{0000-0002-0615-4399}}
\author[2,3,4]{Mick Mulder\,\orcidlink{0000-0001-6867-8166}}
\author[3,5]{Maarten van Veghel\,\orcidlink{0000-0001-6178-6623}}

\affil[1]{Physics Department, University of Lancaster, Lancaster, United Kingdom}
\affil[2]{Van Swinderen Institute, University of Groningen, Groningen, the Netherlands}
\affil[3]{Nikhef National Institute for Subatomic Physics, Amsterdam, the Netherlands}
\affil[4]{Fakultät Physik, Technische Universität Dortmund, Dortmund, Germany}
\affil[5]{Gravitational Waves and Fundamental Physics, Maastricht University, Maastricht, the Netherlands}

\date{}

\begin{document}

\maketitle
\thispagestyle{firstpage}

\begin{abstract}
	These proceedings present the outcome of a feasibility study using RapidSim simulation software that demonstrates that the LHCb experiment will be capable of observing the decays \Bptotaunu and \Bctotaunu using the data that is being collecting during Run 3 of the LHC. The proposed analysis exploits the small distance of only 5.1 millimetres between the sensing elements of LHCb’s innermost silicon pixel detector, the VELO, and the LHC’s proton beams to identify direct pixel hits in the VELO that can be associated with the charged $B^+$, $B_c^+$ or $\tau^+$ particles. By using this extra information, the limitations due to the missing momentum and vertex information will be significantly reduced. This provides enough statistical power to pursue the measurements of these two decay channels at the LHC. In particular for the decay \Bctotaunu, which has been identified by the high energy physics community as a key objective for experiments at the planned next-generation particle accelerators, this means we do not need to wait for the 2030s or beyond to get first experimental constraints.
\end{abstract}

\section{Introduction}
The purely leptonic \Bptotaunu and \Bctotaunu decays are sensitive probes to physics Beyond the Standard Model (BSM)~\cite{Dingfelder:2016twb}, with their SM branching fractions predicted to be~\cite{CKMfitter:2023, Amhis:2021cfy}:
\begin{equation}
\mathcal{BR}(\Bptotaunu)|_{\text{SM}} = \left(0.869^{+0.031}_{-0.030}\right) \times 10^{-4};
\end{equation}
\begin{equation}
\mathcal{BR}(\Bctotaunu)|_{\text{SM}} = (1.95 \pm 0.09) \times 10^{-2}.
\end{equation}

The experimental world average\footnote{Note that this average does not yet include the recent measurement from the Belle II collaboration~\cite{Belle-II:2025ruy}.} of $\mathcal{BR}(\Bptotaunu)$, compiled by the Particle Data Group~\cite{PDG}, is
\begin{equation}
\mathcal{BR}(\Bptotaunu) = (1.09 \pm 0.24) \times 10^{-4},
\end{equation}
which is compatible with the Standard Model prediction, but no discovery has been claimed yet.
In contrast, the decay \Bctotaunu has not yet been experimentally measured. Its first determination is widely recognised as a key milestone in flavour physics and it has been listed among the physics goals of several proposed future facilities, including the Future Circular Collider (FCC)~\cite{Amhis:2021cfy, Zuo:2023dzn, Monteil:2021ith, Krammer:2013vaa} and the Circular Electron Positron Collider (CEPC)~\cite{Zheng:2020ult}.

The interest in \Bctotaunu is further strengthened by the long-standing anomalies observed in the ratios
\begin{equation}
R(D^{(*)}) = 
\frac{\mathcal{BR}(B^0 \to D^{(*)-} \tau^+ \nu_\tau)}{\mathcal{BR}(B^0 \to D^{(*)-} \ell^+ \nu_\ell)}, \qquad \ell \in \{e,\mu\},
\end{equation}
which compare the rates of semileptonic $B$ decays involving a $\tau$ lepton to those involving lighter charged leptons. The current global average of $R(D^{(*)})$ measurements from BaBar, Belle, LHCb and Belle II exhibits a combined tension of about $3.8\sigma$ with respect to the SM expectation~\cite{hflav2025}.

\begin{figure}[htb!]
    \centering
    \begin{subfigure}[t]{0.48\textwidth}
        \centering
        \begin{tikzpicture}
\begin{feynman}
    \vertex (bc);
    \vertex [above left=of bc] (b) {$\bar{b}$};
    \vertex [below left=of bc] (c) {${c}$};
    \vertex [right=of bc] (taunu);
    \vertex [below right=of taunu] (tau) {$\tau^+$};
    \vertex [above right=of taunu] (nu) {${\nu}_{\tau}$};
    
    \diagram* {
        (c) -- [fermion] (bc) -- [fermion] (b),
        (bc) -- [photon, edge label={\(W^+\)}] (taunu),
        (tau) -- [fermion] (taunu) -- [fermion] (nu)
    };
\end{feynman}
\end{tikzpicture}
    \end{subfigure}
    \hfill
    \begin{subfigure}[t]{0.48\textwidth}
        \centering
        \begin{tikzpicture}
  \begin{feynman}
    \vertex (a) ;
    \vertex [left=of a] (f1) {$\bar{b}$};
    \vertex [right=of a] (f2) {$\bar{c}$};
    \vertex at ($(f1) + (0cm, -1cm)$) (f3) {$d$};
    \vertex at ($(f2) + (0cm, -1cm)$) (f4) {$d$};
    \vertex at ($(a) + (1cm, 1.5cm)$) (b);
    \vertex at ($(b) + (1cm, 0.8cm)$) (f5) {$\tau^+$};
    \vertex at ($(b) + (1cm, -0.8cm)$) (f6) {$\nu_\tau$};
    \diagram* {
        (f2) -- [fermion] (a) -- [fermion] (f1),
        (f3) -- [fermion] (f4),
        (a) -- [photon, edge label=\(W^{+}\)] (b) ,
        (f5) -- [fermion] (b) -- [fermion] (f6),

        };

    \draw [decoration={brace}, decorate] (f3.south west) -- (f1.north west) node [pos=0.5, left] {$B^0$};
    \draw [decoration={brace, mirror}, decorate] (f4.south east) -- (f2.north east) node [pos=0.5, right] {$D^{(*)-}$};
    
  \end{feynman}
\end{tikzpicture}
        
    \end{subfigure}
    \caption{Leading-order Feynman diagrams for the purely leptonic decay \Bctotaunu (left) and the semileptonic decay $B^0 \to D^{(*)-} \tau^+ \nu_\tau$ (right).}
    \label{fig:feynmandiagrams}
\end{figure}
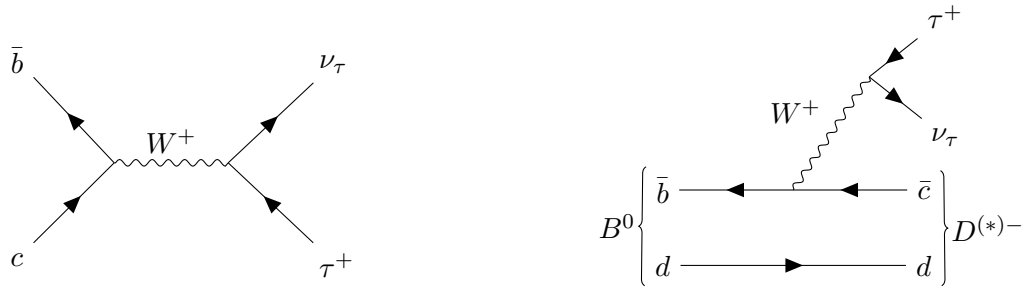

As it can be seen by comparing the SM Feynman diagrams in \Cref{fig:feynmandiagrams}, the $\Bctotaunu$ decay proceeds via the same $b \to c \tau \nu$ quark-level transition that also governs $B^0 \to D^{(*)-} \tau^+ \nu_\tau$, and therefore provides valuable complementary information about the underlying physics processes \cite{Blanke:2018yud,Blanke:2019qrx,Fleischer:2021yjo}.

Observing the decays \Bptotaunu and \Bctotaunu at a hadron collider is considered to be difficult. Because both the $B_{(c)}$ and the subsequent $\tau$ decay produce neutrinos, key kinematic information is lost.
In the busy environment of proton--proton collisions, this makes it challenging to distinguish signal from backgrounds.

The proposed strategy at LHCb is to look for evidence of a charged particle that travelled between the primary vertex (PV) created by the proton--proton collision and the decay vertex of the $\tau$ lepton (TV)~\cite{LHCB-FIGURE-2025-023}. The latter can be reconstructed for \tautothreepi decays, which is the final state considered in this feasibility study.

\begin{figure}[htb!]   
\centering
\resizebox{0.55\textwidth}{!}{%
    \begin{tikzpicture}[scale=0.95]

\newcommand\radius{0.7}

\def\angSVTV{37}
\def\anglePiA{15}
\def\anglePiB{25}
\def\anglePiC{35}
\def\angleNuT{1}
\def\lenpi{4}
  
\coordinate (BL) at (-1,0);
\coordinate (BR) at (10,0);
\coordinate (PV) at ( 0,0);
\coordinate (SV) at ( 3, 1.5);
\coordinate (TV) at (5.5, 3);
\coordinate (NV) at (8.5, 2);

\draw[gray,opacity=0.4,dashed]
(TV);
\draw[-, black,line width=1]
(SV) -> (TV) node[midway, above=0.1,black] {$\tau^+$};

\draw[->, dashed, gray,line width=1]
(SV) -> (NV) node[right,gray] {$\nu_\tau$};

\draw[->,black,line width=1]
(PV) -- (SV) node[midway, above left=0., black] {$B_{(c)}^+$};%, $D_{(s)}^+$};

\draw[->,line width=1] (TV) -- ++(\anglePiA:\lenpi) node[right] {$\pi^+$};
\draw[->,line width=1] (TV) -- ++(\anglePiB:\lenpi) node[right] {$\pi^-$};
\draw[->,line width=1] (TV) -- ++(\anglePiC:\lenpi) node[right] {$\pi^+$};
\draw[->,line width=1, dashed, gray] (TV) -- ++(\angleNuT:3) node[right] {$\bar{\nu}_\tau$};

\fill[black]
(PV) ellipse (2pt and 2pt)
node[black,below=0.3] {\textit{PV}}; 
\begin{comment}
\fill[black]
(SV) circle (2pt) node[below=0.3] {\textit{SV}};
\end{comment}
\fill[black]
(TV) circle (2pt) node[below=0.2] {\textit{TV}};

%hit
% Named paths for B and tau trajectories
\path [name path=pathB] (PV) -- (SV);
\path [name path=pathTau] (SV) -- (TV);
\path [name path=pi1] (TV) -- ++(\anglePiA:5);
\path [name path=pi2] (TV) -- ++(\anglePiB:5);
\path [name path=pi3] (TV) -- ++(\anglePiC:5);

% Loop over VELO planes and check intersections with both
\foreach \x [count=\i] in {2, 3.5, 5, 6.5, 8} {
    \coordinate (VELO\i) at (\x,0);
    \draw[line width=0.5mm, gray, opacity=0.5] (\x,\radius) -- (\x,\radius + 3.5);

    % Define vertical VELO plane path
    \path [name path=plane\i] (\x, -1) -- (\x, 4.5);

    % Draw hits from B and tau paths (red X)
    \ifdim\x pt <2.5pt
    \path [name intersections={of=pathB and plane\i, by=hitB\i}];
    \draw[red, line width=1.5pt, mark=x, mark options={color=red, scale=2.5}]
        plot coordinates {(hitB\i)};
    \fi
    \ifdim\x pt <5.5pt
    \path [name intersections={of=pathTau and plane\i, by=hitTau\i}];
    \draw[red, line width=1.5, mark=x, mark options={color=red, scale=2.5}]
        plot coordinates {(hitTau\i)};
    \fi

                                            }

\end{tikzpicture}
}
\caption{Schematic illustration of a $B_{(c)}^+$ meson decaying to a $\tau^+$ lepton, which subsequently decays to three pions. 
The gray vertical lines represent the VELO sensor modules, with red crosses indicating hits between PV and TV.
The dashed lines represent neutrinos, which are not detected in the LHCb experiment.}
\label{fig:cylinder}
\end{figure}
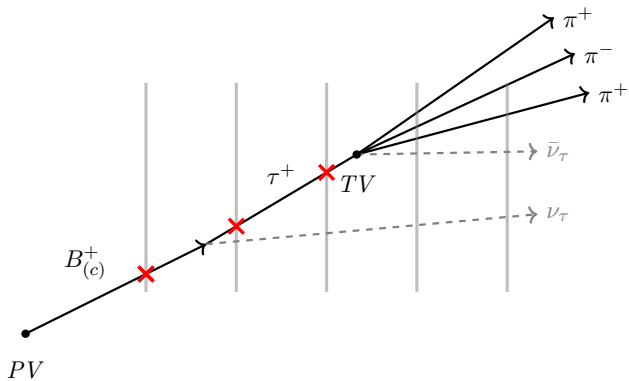

Both the $B_{(c)}$ meson and the $\tau$ lepton are charged, have lifetimes of the order of 1~ps, and therefore travel distances of the order of a centimetre at LHC energies.
Given that the silicon sensors of the upgraded LHCb VELO are positioned as close as 5.1~mm from the LHC beams~\cite{LHCb-TDR-013,Akiba:2024}, there is a non-negligible probability that the $B_{(c)}$ or the $\tau$ traverse a sensor before decaying. This leaves a measurable signal in the VELO, which will be referred to as \textit{VELO-hit} (\Cref{fig:cylinder}).

Selecting events containing at least one VELO-hit places strong constraints on the minimum flight distance between the PV and the TV. 
Furthermore, the location of the VELO-hit closest to the PV provides a more accurate estimate of the initial flight direction of the $B$ meson than the line connecting the PV and the reconstructed $\tau$ vertex. 
This improved directional information enhances the kinematic reconstruction of the event and strengthens the separation of signal from background.

%The remainder of the proceedings presents the simulation of the relevant $B$-meson decays and the detector response, followed by the results of the feasibility and sensitivity study for the \Bptotaunu and \Bctotaunu decays, concluding with a summary of the main findings.

\section{Data simulation}\label{sec:sim}
The feasibility study is based on RapidSim~\cite{RapidSim, Cowan:2016tnm}, a tool designed for fast simulations of heavy-flavour hadron decays.
RapidSim provides the kinematics of the decays of interest, but does not model proton--proton collisions or interactions with the detectors.

Background events with a similar experimental signature to that of \Bptotaunu and \mbox{\Bctotaunu}, i.e.~three charged pions originating from a common displaced vertex, can originate from charm- and beauty-hadron decays.
We include the following three background cocktails, matching the list of decay channels used in the feasibility study for FCC-ee~\cite{Amhis:2021cfy}:

\begin{itemize}
    \item \textbf{$D\rightarrow\tau\nu$}, consisting of $D^+\to\tau^+\nu_{\tau}$ and $D_s^+\to\tau^+\nu_{\tau}$;
    \item \textbf{$B\rightarrow D\,3\pi$}, consisting of $B_h\rightarrow D_h\pi^+\pi^+\pi^-$ and $B_h\rightarrow D_h^*\pi^+\pi^+\pi^-$;
    \item \textbf{$B\rightarrow DY$}, consisting of $B_h\rightarrow D_h\tau^+\nu_{\tau}$, $B_h\rightarrow D_h^*\tau^+\nu_{\tau}$, $B_h\to D_h D_s^+$, \mbox{$B_h\to D_h^*D_s^+$} and $B_h\to D_h^*D_s^{*+}$.
\end{itemize}
Here, $B_h$ denotes any of $\{B^+, B_d^0, B_s^0, \Lambda_b^0\}$, with the corresponding $D_h$ being one of $\{\bar D^0, D^-, D_s^-, \Lambda_c^-\}$. The intermediate decays $D_s^{*+} \to D_s^+ \gamma$,  $D_s^+ \to \tau^+ \nu_\tau$, and \tautothreepi, are implicitly assumed. In this study, $D_h$ is assumed to not contribute to the reconstructed final-state particles and it is therefore ignored within RapidSim.

%\subsection{Candidate selection}

Signal and background events are first filtered based on the geometric acceptance of the LHCb detector and the application of the following selection cuts:
\begin{itemize}
    \item the invariant mass of the two combinations of oppositely charged pions, $m_{\pi^+ \pi^-}$, satisfies \mbox{$m_{\pi^+ \pi^-} \le 1670 \text{ MeV}$}, which roughly corresponds to $m_{\pi^+ \pi^-} \le m_{\tau}-m_{\pi}$;
    \item the invariant mass of the three pions, $m_{3\pi}$, satisfies $500\text{ MeV} \le m_{3\pi} \le 1800 \text{ MeV}$;%, which suppresses contributions from low-mass decays;
    \item the transverse momentum of the three pion system exceeds 5 GeV;
    \item at least one VELO-hit is present between the PV and TV, confirming the existence of a charged intermediate particle.
\end{itemize}

A factor $\varepsilon_{\rm iso}$ is applied to the event yields of $B \to D 3\pi$ and $B \to D Y$ decays, where additional detectable particles beyond the three signal pions are present. $\varepsilon_{\rm iso}$ represents the probability that all these additional particles escape reconstruction, causing the background event to mimic the three-pion signal topology. 
A conservative value of $\varepsilon_{\rm iso} = 10\%$ is chosen based on both performance numbers from existing tools at LHCb~\cite{Calvi:2025rme} and reconstruction efficiencies of both charged and neutral particles~\cite{LHCb-DP-2014-002, LHCb:2014nio, LHCb-2003-091,Sobczak:2011dva}.

The expected number of events is calculated by combining the relevant production cross sections, the branching fractions of the decays, and the integrated luminosity of the dataset, weighted for an efficiency factor combining the geometric acceptance of the detector, the applied selection cuts, and the probability that additional particles escape reconstruction.

\section{Sensitivity study}\label{sec:sens}
A sensitivity study is performed to determine the feasibility of observing the \Bptotaunu and \Bctotaunu decays at the LHCb experiment using the data collected during Run 3 of the LHC. The study is based on 2000 pseudo-experiments, in which the signal yields are extracted from an extended two-dimensional binned maximum-likelihood fit to toy datasets. The two fitted observables are the \emph{corrected mass} and the output of a boosted decision tree (BDT) classifier.

The corrected mass is defined as
\begin{equation}\label{eq:mcorr}
    m_{corr} = \sqrt{m_{3\pi}^2 + |\vec p^{\perp}_{3\pi}|^2} + |\vec p^{\perp}_{3\pi}|,
\end{equation}
where $\vec p^{\perp}_{3\pi}$ is the momentum of the three-pion system transverse to the $B$-meson flight direction, approximated by the line connecting the PV to the first VELO-hit.

The BDT is trained to discriminate \Bctotaunu signal from background using the distributions of the three-pion kinematic and topological observables, including momentum, impact parameter, flight distance, and invariant mass.

\begin{figure}[htp!]
    \centering
    \includegraphics[width=0.48\textwidth]{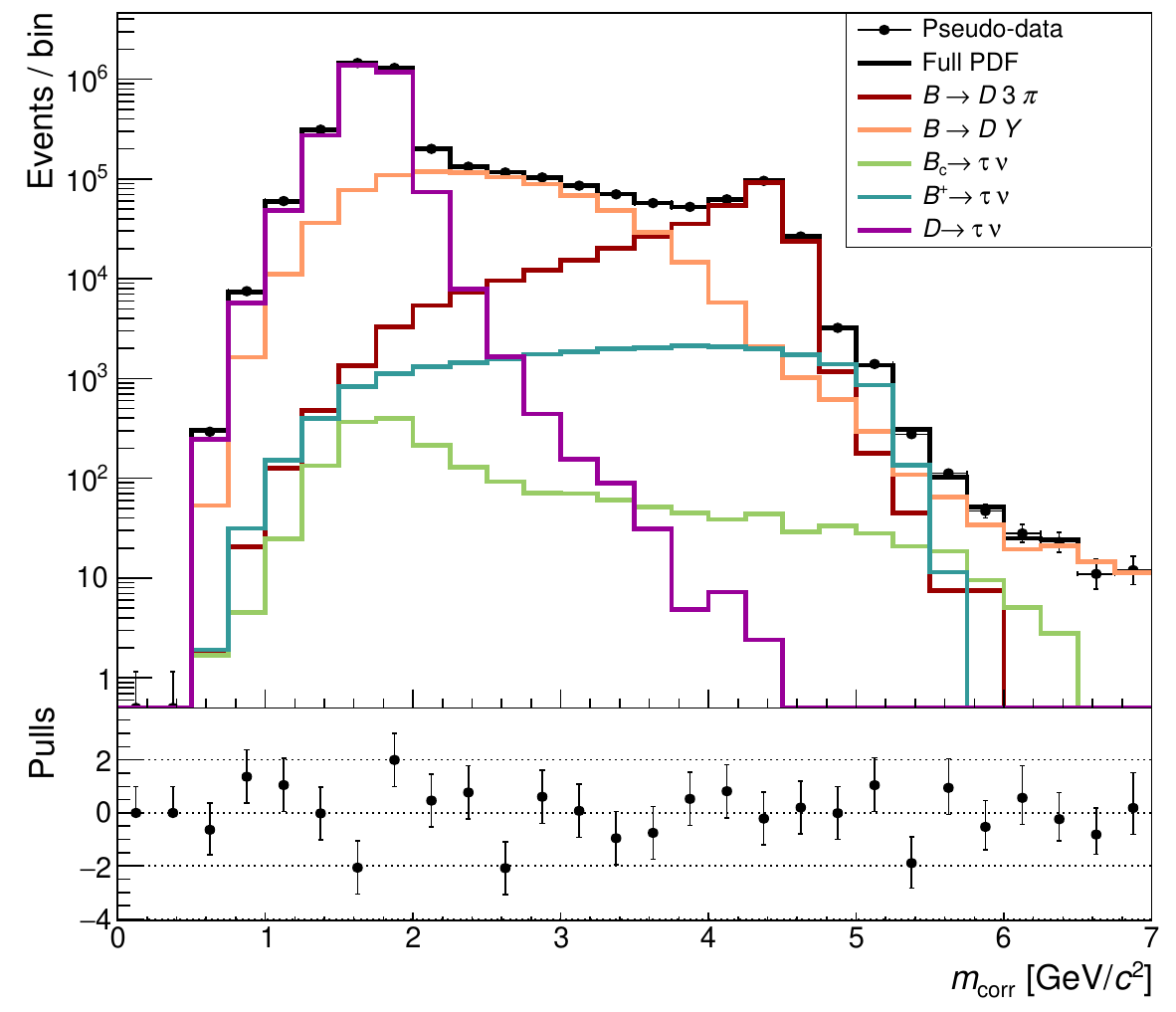}
    %\hfill
    \includegraphics[width=0.48\textwidth]{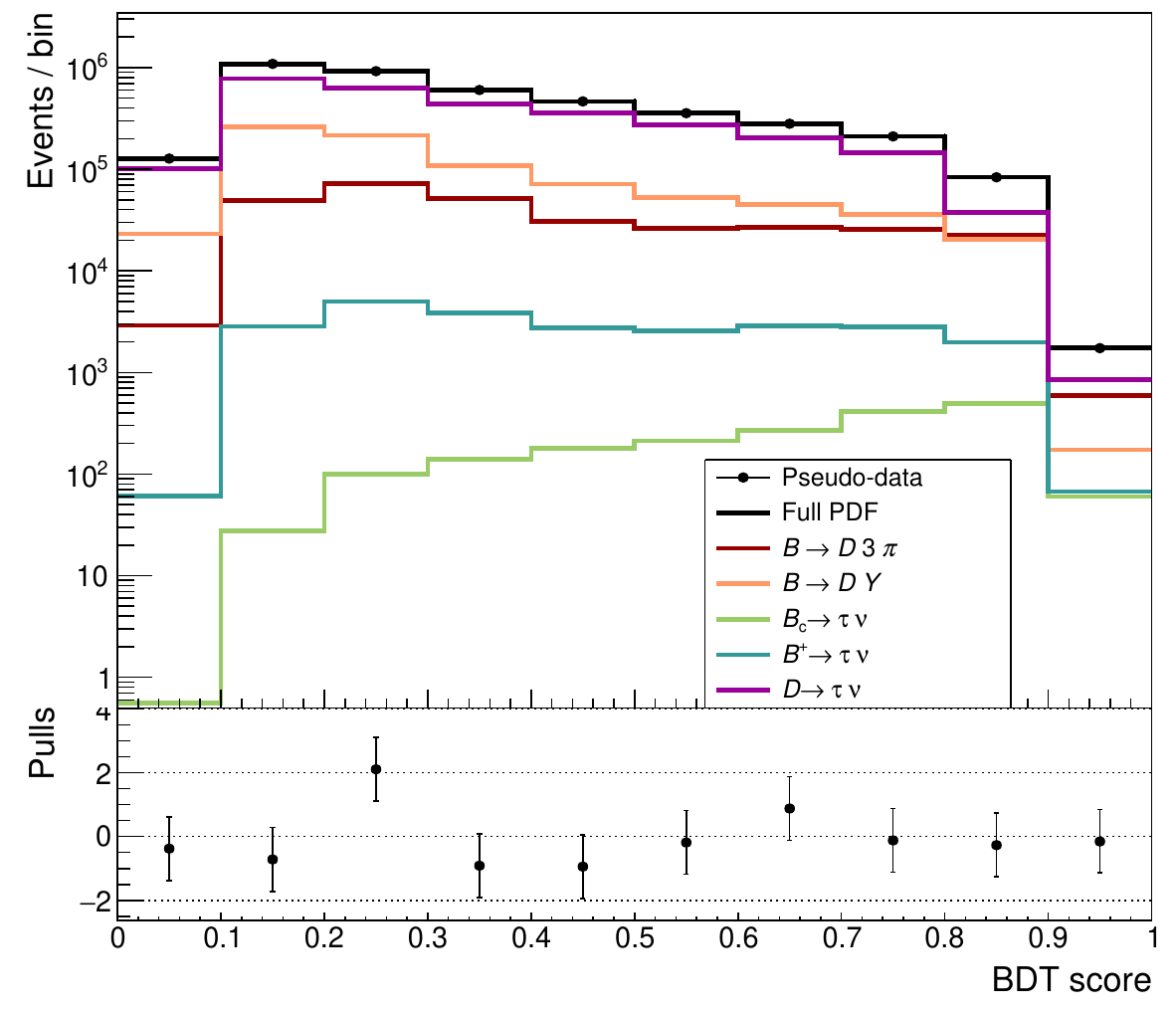}
    \caption{Distributions of $m_{corr}$ (left) and the BDT classifier output (right), showing the projections of the two-dimensional fit on a representative pseudo-dataset generated with an integrated luminosity of $10\,\mathrm{fb}^{-1}$.}
    \label{fig:fit_projections}
\end{figure}

In the fit, the template shapes of all five components (\Bptotaunu, \Bctotaunu, $B\to D\,3\pi$, $B \to D\,Y$ and $D\to \tau \nu$) are fixed, with the event yields being the only free parameters.

\begin{figure}[htb!]
    \centering
    \includegraphics[width=0.9\textwidth]{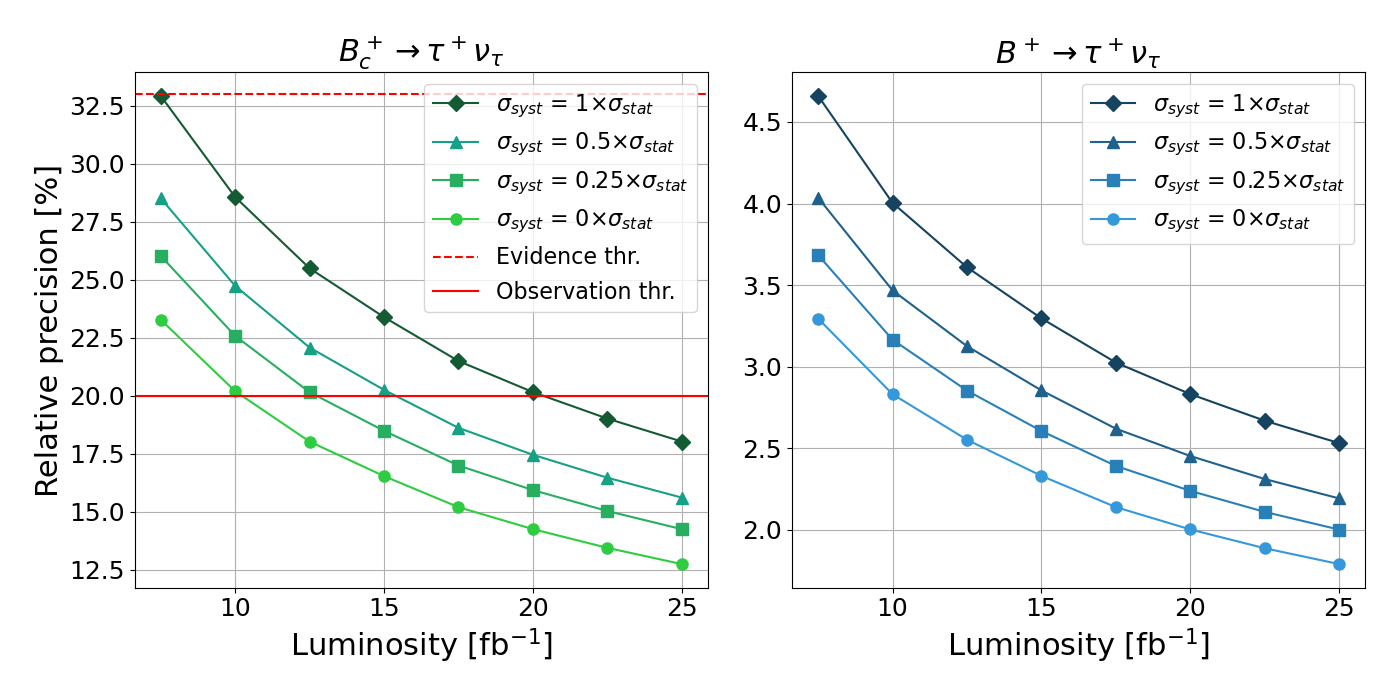}
    \caption{Mean relative precision on the fitted \Bctotaunu (left) and \Bptotaunu (right) yields as a function of integrated luminosity, based on 2000 pseudo-experiments per point. Different curves correspond to scenarios where the systematic uncertainty is assumed to be 0, 25\%, 50\%, or 100\% of the statistical uncertainty. These scenarios are included to illustrate the impact of potential systematic effects not explicitly evaluated in this study. For the \Bctotaunu decay, the dashed red line indicates the $3\sigma$ significance level threshold, while the solid red line corresponds to $5\sigma$.}\label{fig:MCresults}
\end{figure}

The projections of a representative fit performed with an integrated luminosity of $10~\mathrm{fb}^{-1}$ onto $m_{\text{corr}}$ and the BDT score are shown in \Cref{fig:fit_projections}.
The fit procedure is repeated for several luminosity scenarios. For each scenario, the mean relative uncertainty on the fitted \Bptotaunu and \Bctotaunu yields is defined as the average, over the 2000 pseudo-experiments, of the ratio $\sigma_{\text{fit}}/n_{\text{fit}}$, where $n_{\text{fit}}$ is the fitted yield and $\sigma_{\text{fit}}$ its associated uncertainty.
The resulting relative uncertainties are plotted in \Cref{fig:MCresults} as a function of the integrated luminosity.  
The figure also illustrates the impact of different levels of systematic uncertainty, each expressed as a fraction of the statistical uncertainty, on the overall precision of the measurement.

\section{Results and conclusions}\label{sec:con}
Based on the presented feasibility study, the observation of the \Bptotaunu and \Bctotaunu decays at the LHCb experiment appears to be within reach using data from LHC Run~3.  

Considering only statistical uncertainties, the \Bctotaunu decay can be observed with a data sample corresponding to an integrated luminosity slightly above 10~fb$^{-1}$.  
When including a systematic uncertainty of the same size as the statistical uncertainty, a dataset of slightly more than 20~fb$^{-1}$ would be required to claim the first observation.  
Given the luminosity already collected by LHCb since 2024~\cite{LHCb-OperationsPlots}, such a dataset is expected to be available by mid-2026.  

For the \Bptotaunu decay, a relative precision smaller than 5\% is obtained for all considered luminosity scenarios, indicating that the observation of this decay is already feasible with the data collected to date.

These results demonstrate that LHCb has the potential to provide the first experimental constraints on \Bctotaunu and a precise measurement of \Bptotaunu, offering complementary information to the ongoing tests of lepton flavour universality in $R(D^{(*)})$ and the search for physics beyond the Standard Model.

\bibliographystyle{ieeetr}
\bibliography{references}

\end{document}